\documentclass[fleqn,usenatbib]{mnras}

\usepackage[T1]{fontenc}

\DeclareRobustCommand{\VAN}[3]{#2}
\let\VANthebibliography\thebibliography
\def\thebibliography{\DeclareRobustCommand{\VAN}[3]{##3}\VANthebibliography}

\usepackage{graphicx}	
\usepackage{amsmath}	
\usepackage{amssymb}	
\usepackage{pdflscape}  

\title[The initial size of SG in GCs]{Monte Carlo simulations of multiple populations in globular clusters: constraints on the initial size of the second generation from binary stars}

\author[A. Sollima et al.]{
A. Sollima,$^{1}$\thanks{E-mail: antonio.sollima@inaf.it (AS)}
R. Gratton,$^{2}$
S. Lucatello,$^{2}$
E. Carretta,$^{1}$
\\
$^{1}$INAF Osservatorio di Astrofisica e Scienza dello spazio di Bologna, 
via Gobetti 93/3, 40129 Bologna, Italy
\\
$^{2}$INAF Osservatorio Astronomico di Padova, vicolo dell'Osservatorio 5, 35122 Padova, Italy
}

\date{Accepted 2022 February 14. Received 2022 February 8; in original form 2021 November 3}

\pubyear{2021}

\begin{document}
\label{firstpage}
\pagerange{\pageref{firstpage}--\pageref{lastpage}}
\maketitle

\begin{abstract}
We present the result of a survey of Monte Carlo simulations of globular clusters hosting two generations of stars including a large ($f_{b}=50\%$) fraction of primordial binaries in both populations.
The dynamical evolution of the two stellar populations is followed for a Hubble time taking into account the effect of the tidal field, two-body relaxation, stellar evolution and three/four-bodies interactions.
The fraction of surviving binaries, once accounted for the observational bias and uncertainties, is compared with the available radial velocity time-series performed in real globular clusters, and it is used to constrain the initial spatial concentration of the second generation.
The fraction of second generation binaries appears to depend only on the ratio between the total cluster mass and the initial size of the second generation which determines the average velocity dispersion across the extent of this stellar population. 
In spite of the various uncertainties, we find that the observed fraction can be obtained only assuming a strong initial concentration of the second generation ($r_{h,S}\sim 0.1~(M/10^{6} M_{\odot}) pc$). 
The evolution of the first generation binary fraction is more sensitive to the tidal field strength (with a non negligible effect of the cluster orbital eccentricity) since the tidal field has a direct impact on the first generation structural properties.
\end{abstract}




\section{Introduction}
\label{intro_sec}

The existence of multiple populations in globular clusters (GCs) is a commonly accepted paradigm today.
In particular, two or more stellar populations characterized by different abundances of p-capture elements have been detected in almost all Galactic GCs \citep{carretta2009,milone2017}.
This evidence implies that GCs were able to retain part of the gas expelled from the most massive stars of their oldest population (hereafter referred as "first generation"; FG) and to recycle it to form the youngest ones (hereafter "second generation"; SG).

In recent years, many authors proposed various models of GC formations aimed at reproduce the whole body of observational evidence \citep[see][for recent reviews]{bastian2018,gratton2019}.
A class of these models assumes an extended FG whose ejecta are recycled to form the SG in its central region \citep[the so-called "multiple-generations" scenario; ][]{decressin2007,dercole2008,krause2013}, while others predict a compact FG polluting a fraction of its stars in its central region \citep{demink2009,bastian2013,gieles2018}. 
So, while different models predict different initial sizes and masses for the FG, they all converge in predicting a concentrated SG. 

Indeed, despite a general consensus on the mechanism of formation and chemical pollution of multiple populations is far from being reached, there are a few pieces of evidence suggesting a compact initial structure of the SG.
Specifically, SG stars \citep[constituting $\sim60\%$ of the whole cluster population; ][]{carretta2009,milone2017} are observed today systematically more concentrated than FG in many GCs \citep{lardo2011,dalessandro2019}, they host a lower fraction of binaries \citep{dorazi2010,lucatello2015,milone2020} and show in a few cases a higher degree of radial anisotropy \citep{richer2013,bellini2018,libralato2019}.
Since there is no known dynamical mechanism able to create {\it ex-novo} a differential radial segregation of FG/SG stars, the above differences must be linked to primordial structural differences.
In particular, all these pieces of evidence are consistent with a scenario in which the SG forms in the central region of the proto-cluster characterized by a high density and kinetic temperature.
In such an environment, two-, three- and four-body interactions are frequent and lead to an efficient destruction of binaries and to a diffusion of SG stars outwards on radial orbits \citep{henault2015,tiongco2019}.
Unfortunately, the subsequent dynamical evolution progressively mixes the two populations and the leftover of the differences could be visible today only in less dynamically evolved GC \citep[i.e. those characterized by a long relaxation time; see e.g. ][]{decressin2008,dercole2008,vesperini2013,vesperini2018,vesperini2021,sollima2021}. 
Because of the mixing effect of two-body relaxation, it is extremely hard to put quantitative constraints on the initial concentration of SG stars even for the GCs in which two-body relaxation has not completely erased the differences in the spatial distributions of the two stellar generations.
Indeed, the efficiency of two-body relaxation depends on quantities (mass, half-mass radius, mass-function) which vary during the cluster evolution and are affected by the interaction with the tidal field so that it is impossible to trace back the initial conditions from the knowledge of present-day structure.

A more promising approach relies on the analysis of the fraction of binaries in the different populations.
Binaries are indeed destroyed as a result of close collisions with singles and other binaries.
The efficiency of the binary destruction process depends on the frequency and kinetic energy of collisions, being larger in denser regions or in those characterized by a larger velocity dispersion \citep{hut1983} i.e. in the central region of the cluster.
So, because of the initial concentration of the SG, a preferential destruction of SG binaries occurs at early epochs.
During the subsequent evolution, the two populations tend to mix, reducing their differences in density, size and, consequently, binary destruction rate without significantly altering the ratio of FG/SG binary fractions.
While the above process is not expected to produce noticeable differences in the binary fractions of FG/SG at the extremes of the binaries binding energy distribution 
(where very-hard and very-soft binaries will entirely survive/ionize in both populations), 
there is a range at intermediate binding energies where striking differences are expected.
These differences therefore reflect the instant of maximum contrast between the sizes of the two stellar populations occurring at their formation.

Observationally, the most extensive survey of binaries able to distinguish FG/SG stars has been performed by \citet{lucatello2015}.
From their study, based on radial velocity monitoring in 10 GCs along a time baseline $>3~yr$, they derived fractions of binaries $f_{b,F}=4.9\%\pm 1.3\%$ and $f_{b,S}=1.2\%\pm 0.4\%$ for the FG and the SG, respectively.
Note that, because of their observational cadence, this analysis is sensitive to short-period ($P<100~d$) binaries.
The subsequent works of \citet{dalessandro2018} and \citet{kamann2020} on individual GCs confirmed this result.
The independent analysis of \citet{milone2020}, based on photometric detection of binaries in 4 GCs, found a similar binary fraction among FG/SG in the central region probed by their data in 3 out 4 GCs confirming the excess of FG binaries in the remaining one.
The discrepancy between spectroscopic and photometric binary fractions can be explained considering the different radial extent of the two analyses: 
while the work of \citet{lucatello2015} observe data over a wide area around the half-light radius, the HST data of \citet{milone2020} only cover the central portion of GCs where the decrease due to binary destruction is partially compensated by their increase as a result of mass segregation \citep{hong2019}.

From the theoretical side, the first attempt to determine the differential effect of binary destruction in FG/SG has been made by \citet{vesperini2011}.
By using a hybrid N-body/analytic approach, these authors estimated that the destruction rate of SG binaries can be larger up to an order of magnitude than that of FG, 
in particular when long-period binaries or large initial FG/SG radial ratios are considered.
N-body simulations focussed on the evolution of soft and hard binaries have been performed by \citet{hong2015,hong2016}.
Also in this case, the preferential destruction of SG binaries has been confirmed. 
All the above studies, while investigating a wide range of initial conditions, cannot be directly compared with observations because they are limited to binaries with a specific semi-major axis or binding energy.
In their review, \citet{gratton2019} adopted an extremely simplified analytic model that,
neglecting dynamical evolution, was used to predict the expected ratio between FG/SG binary fractions.
They showed that the ratio of FG/SG binary fractions is a unique function of the initial FG/SG half-mass radii ratio, 
and binary fractions consistent with those measured by \citet{lucatello2015} can be obtained only assuming an initial ratio $r_{h,S}/r_{h,F}<0.2$. 
Unfortunately, the simplified framework adopted by these authors limits the application of such an approach to determine a strong constraint on the initial condition of the SG. 

In this paper, we use Monte Carlo simulations of stellar systems hosting two stellar populations containing a large fraction ($f_{b}=50\%$\footnote{We define the binary fraction as the ratio between the number of binaries and the total number of systems (singles+binaries).}) of binaries and spanning a wide range of initial conditions affecting the evolution of the binary fractions.
The fraction of surviving binaries in the last snapshot of these simulations is compared with observations and used to derive a constraint on the initial half-mass radius of the SG.

In Sect. \ref{code_sec} the Monte Carlo code adopted for the simulations is presented. 
In Sect. \ref{setup_sec} the initial setup of simulations is described.
Sect. \ref{res_sec} is devoted to the presentation of the 
results and the comparison with observational data.
We summarize our results in Sect. \ref{concl_sec}.

\section{Monte Carlo code}
\label{code_sec}

The simulations have been run using the Monte Carlo code originally presented in \citet{sollima2014}.
It follows the \citet{henon1971} scheme to simulate the dynamical evolution of a stellar system accounting for the effect of two-body relaxation and tidal effects in various kinds of Galactic potentials.
In this approach, at each timestep the integrals of motions (energy and angular momentum) of particles are perturbed assuming interactions between stars with a contiguous ranking in distance from the cluster centre, and random impact parameters.
The shape of the potential is updated accordingly and a new statistical realization of the particle distribution is performed by placing particles randomly across their orbit.
So, at odds with N-body simulations where a timestep smaller than the dynamical time ($t_{dyn}$) must be adopted to accurately integrate the orbits of particles, in the Monte Carlo approach the timestep must be smaller than the relaxation time ($t_{rel}>>t_{dyn}$). 
This technique has been proved to be effective in simulating the evolution of stellar systems, providing comparable results of N-body simulations with a striking gain in computing time \citep{giersz2013,rodriguez2016}. 
Such an increased efficiency allows to run extended sets of simulations of stellar systems with masses comparable to those of present-day GCs \citep{kremer2020}.
In recent years, several groups have developed efficient Monte Carlo codes adding additional levels of complexity to account for 
the presence of a mass spectrum, stellar evolution, three- and four-body interactions and the interaction with a tidal field 
\citep{giersz1998,joshi2000,sollima2014,vasiliev2015}.

Binaries, the focus of this work, can be present at the beginning of the simulations or form during its evolution.
During the cluster evolution they interact between them and with the other single stars leading to a number of possible outcomes (ionization, exchange, fly-by, triples, etc.).
At each timestep, the probability of a close encounter between two or three neighbor objects (either single-single, single-binary, binary-binary or 3 singles) within a sphere of radius $r_{m}$ is calculated \citep[see ][]{sollima2019,sollima2021} and the interaction is integrated using a symplectic algorithm \citep{yoshida1990} if a random number uniformly distributed between 0 and 1 is smaller than such a probability. 
The initial conditions of the interaction are set by randomly rotating the binary rotation plane(s) and placing the interacting objects at a distance such that the potential energy felt by the involved objects is 20 times smaller than the binary binding energy. 
The particles move towards their centre of mass with a relative velocity calculated from their relative velocity at infinity with an impact parameter randomly extracted across the $\pi r_{m}^{2}$ cross section. 
The interaction is followed until the total energy of the system exceeds zero. 
The products of the collision are then placed with their final velocities in the cluster reference frame.
We address the reader to the \citep{sollima2014,sollima2019,sollima2021} papers for a detailed description of the adopted code.

Here we describe only the updates to the code which have been adopted to handle a detailed treatment of stellar evolution.
Stellar evolution in single stars has been implemented adopting the set of tracks of \citet{pietrinferni2006} and \citet{limongi2018}.
These models include all the most updated physics (e.g. winds, angular momentum losses and pulsation for massive stars), providing the 
stellar radii, core masses, surface temperature and luminosities as a function of mass and time.
The stellar lifetimes have been derived by interpolating through the stellar tracks at various metallicities and $\alpha-$enhancements.
After this timescale, stars finish their evolution leaving a compact remnant according to their original masses.
Low-mass ($M<8~M_{\odot}$ at solar metallicity) stars leave a white dwarf.
Stars developing a non-degenerate CO core ($M>8~M_{\odot}$ at solar metallicity) undergo a fast evolution ending with a SNe II explosion.
The remnant left by these stars is either a neutron star (NS; if $8<M/M_{\odot}<30$ at $[Fe/H]=0$) or a black hole (BH; for more massive stars).
For these stars a velocity kick has been added to mimic the effect of SNeII explosion.
In particular, the kick velocity has been extracted from a Maxwellian distribution with $\sigma_{kick}=100~km/s$ for stars leaving a neutron star remnant ($8<M/M_{\odot}<30$ at $[Fe/H]=0$) and $\sigma_{kick}=80~km/s$ for those leaving a black hole remnant \citep{kruijssen2009}.
The initial-final mass relation of all objects has been taken from the relations of \citet{kruijssen2009}.
The radii of dark remnants have been adopted from the mass-radius relations of \citet{hurley2000}. 
More details of the implementation of stellar evolution for single stars in our Monte Carlo code are described in \citet{sollima2021}.

When stellar evolution proceeds in the components of a binary star several processes like mass transfer, coalescence, common envelope evolution, etc. can occur.
The criterion adopted for Roche-lobe overflow is that the stellar radius of either component exceed the volume-averaged Roche lobe radius \citep[from][]{paczynski1976} calculated at the orbital periastron.  
The amount of mass transferred from the donor and that accreted by the accretor depends on the evolutionary stage of the two stars and have been calculated using the recipes of \citet{hamers2021}.
In binaries with a significant eccentricity, orbit averaged rates have been adopted. 
If at least one of the component of the binary is in a giant phase, its stellar parameters can change on timescales comparable to the simulation timestep ($\Delta t$). 
In this case, the average mass transfer rate has been calculated along the entire post-Main Sequence stage of the evolved star(s) ($\Delta t_{RG}$) and an effective rate $\dot{m}_{eff}=\langle \dot{m}\rangle\Delta t/\Delta t_{RG}$ has been adopted. 
In case of slow mass transfer ($\dot{m}\Delta t<m$), the semi-major axis and eccentricity of the binary have been updated adopting the prescriptions for weak friction of \citet{hut1981}.
Instead, in case of a sudden mass change, the orbit of the binary has been numerically integrated until a new stable configuration is reached.
The effect of mass transfer can lead to a rejuvenation, to an evolutionary phase change or to the creation of exotic objects like Blue Straggler stars, Novae, SNe Ia, X-ray binaries, cataclysmic variables and millisecond pulsars.
All these processes have been modelled using the recipes of \citet{hurley2002}.

As a further improvement, the code has been parallelized using the {\rm openMP} multithreading implementation.
Thanks to the above modification, our Monte Carlo code is now able to run simulations with $N=2\times10^{6}$ particles with large ($f_{b}=50\%$) fraction of binaries (see Sect. \ref{setup_sec}) in $\sim$1 day using a four 3.3GHz Intel processors machine with 4Gb of RAM.

The various improvements to the code need to be tested against N-body simulations.
For this purpose, we run additional simulations with the same setup of three 
different works: {\it i)} the simulations of 
\citet{vesperini2013} simulating the evolution of two nested stellar populations, 
{\it ii)} the simulations of \citet{hong2015} including a significant population of primordial binaries, and
{\it iii)} the simulations of \citet{baumgardt2003} including a mass 
spectrum and the effect of stellar evolution.
The results of such comparisons are show in the Appendix A.

\section{Initial setup}
\label{setup_sec} 

To investigate the effect of the various structural parameters on the fraction of surviving binaries, we performed a set of simulations spanning a range of initial conditions.
As a working hypothesis, we defined our initial setup in the framework of the scenario proposed 
by \citet{dercole2008}. In this model, the SG forms in a time interval between 30 and 100 Myr after the cluster formation from the ejecta of intermediate-mass ($4<M/M_{\odot}<8$) Asympthotic Giant Branch (AGB) stars of the FG which cool and mix 
with a fraction of pristine gas, residual from the formation of the FG. 
Although there is not a general consensus on the validity of this model (see Sect.\ref{intro_sec}), it is useful to define a set of constraints on the initial conditions of the FG (see below).
Note that all models within the "multiple-generations" scenario, predict a formation of the SG on a timescale shorter than that needed to ensure a significant number of interactions involving binaries. 
So, if different polluters are considered \citep[like e.g. the Fast Rotating Massive Stars; ][]{krause2013}, 
the differences in the timing of the onset and the duration of the SG formation burst should not have a significant effect on the
survival of binaries after a Hubble time.

The number of binaries changes during the cluster evolution mainly because of two effects: evaporation and ionization \citep[see e.g.][]{sollima2008}.
Binaries are indeed on average more massive than single stars and lose velocity in the elastic interactions with their neighbours, thus populating low energy orbits and being preferentially retained with respect to single stars.
Close interactions with single stars or other binaries can lead to inelastic encounters where part of the kinetic energy of the colliding object is absorbed by the binary and used to separate its components.
If the amount of transferred energy exceeds the binary binding energy the binary is disrupted \citep{heggie1975}.

In principle, all the parameters affecting the structural evolution of the cluster can alter the efficiency of the above processes and, consequently, the fraction of surviving binaries.
So, a comprehensive survey of initial conditions should explore a parameter space made of FG/SG masses, half-mass radii, concentrations, velocity anisotropies, mass functions, binary fractions and characteristics (period, mass-ratio and eccentricity distributions), and GC systemic orbits.
Of course, such an extended grid of simulations is out of the possibility of any computing facility.
However, part of the above space of parameters can be reduced by eliminating those parameters affecting the cluster evolution at the second order, and using independent constraints to link some of them.

For this reason all the simulations run in this work assume a \citet{kroupa2001} Initial Mass Function (IMF) defined between 0.1 and 120 $M_{\odot}$ for the FG and between 0.1 and 8 $M_{\odot}$ for the SG \citep[to avoid the explosion of SG SNeII; ][]{dercole2008}.

The simulated clusters move in a logarithmic potential with circular velocity $v_{circ}=220~km/s$ following 
circular orbits at various galactocentric distances ($R_{G}=5,~8.5$ and $20~kpc$). 
To test the effect of the adopted tidal field, one simulation has been run adopting 
the tidal field generated by the axisymmetric potential of \citet{johnston1995} and the orbit of the GC NGC1851 
calculated using the systemic distance, radial velocity and proper motions provided by \citet{baumgardt2021} and \citet{vasiliev2021}. This last orbit is extremely eccentric ($e=0.9$) with a 
peri-Galactic distance of $\sim$1 kpc and an orbital period of $\sim$450 Myr, 
so that some 60 bulge- and 120 disk- shocking occur during the entire cluster evolution.

We follow the initial conditions adopted in the simulations run in \citet{sollima2021} in the "multiple-generations" scenario 
which have been shown to lead to a final FG/SG ratio consistent with that observed in real GCs \citep[$N_{SG}/N_{t}>50\%$; ][]{carretta2010,milone2017} assuming that
{\it i)} FG stars are distributed with a strong degree of primordial mass-segregation \citep[but see also][]{vesperini2021}, and 
{\it ii)} the initial Roche-lobe filling factor of the FG is $r_{h,F}/r_{J}>0.15$.
So, we adopt an isotropic distribution of FG stars according a \citet{gunn1979} multimass model with $W_{0}=25$\footnote{We 
adopt the definition of the central adimensional potential $W_{0}=\Phi_{0}/\sigma_{K,1}^{2}$, where $\Phi_{0}$ is the central potential and $\sigma_{K,1}$ is an energy normalization proportional to the central velocity dispersion of the most massive considered stellar group ($M_{1}=120~M_{\odot}$). 
For reference, adopting $W_{0}=25$ the stellar groups at 1, 25 and 50 $M_{\odot}$ follow profiles similar to single-mass \citet{king1966} models with $W_{0}=$3, 4 and 5, respectively.}.
The half-mass radius of FG is set to $r_{h,F}=0.2~r_{J}$, where 
\begin{equation*}
r_{J}=R_{G}^{2/3}\left(\frac{G M_{F}}{2 v_{circ}^{2}}\right)^{1/3}
\label{rj_eq}
\end{equation*}
is the Jacobi radius and $M_{F}$ is the FG initial mass.
Between 30 Myr and 100 Myr after the beginning of the simulation, SG stars are 
continuously added in the central region of the 
cluster following a \citet{king1966} profile with central adimensional 
potential $W_{0}=5$ and a various half-mass radii ($r_{h,S}=0.5,~1$ and $3.5~pc$). 
The energies and angular momenta of SG stars are drawn solving the anisotropic Jeans equation 
using the instantaneous cluster potential and adopting an Osipkov-Merrit radial 
anisotropy profile \citep{osipkov1979,merritt1985} with anisotropy radius equal to $r_{a}=r_{h,S}$. 

The initial FG mass has been set to $M_{F}=2\times 10^{6}~M_{\odot}$. 
This value has been chosen as a compromise to ensure a final mass after a Hubble time which is consistent with that typical of present-day GCs ($\sim10^{5}~M_{\odot}$) still maintaining an affordable number of particles. 
In one simulation We adopt a smaller FG mass of $M_{F}=6\times 10^{5}~M_{\odot}$ to test the impact of the FG mass.
The SG mass has been set on the basis of the results of \citet{dercole2010} who, assuming ad-hoc yields for AGB and super-AGB stars, 
required an amount of pristine gas such that the emerging SG has a mass $\sim$10\% of that of the FG to reproduce the distribution of stars in the Na-O anticorrelation plane.
In one case a SG with a mass only 3\% of the FG mass has been simulated.

The binary population has been simulated by random pairing stars from a 
\citet{kroupa2001} IMF. 
Periods and eccentricities have been extracted from the 
distributions reported in \citet{duquennoy1991}. 
From an initial library, binaries with semi-major axes smaller 
than the limit for stable overflow described in Sect. \ref{code_sec} were removed.
In all the considered simulations the initial fraction of binaries is $f_{b}=50\%$ for both populations, consistent with that found in the Galactic field at low-metallicity \citep{moe2019}.
This choice implicitly assumes that the process of binary formation in GCs (both FG and SG) occurred in the same fashion of what occurred among solar-type stars in the Galactic field, with the observed differences being entirely due to the long-term dynamical evolution. 
The arbitrariness of this assumption is justified by the consideration that, according to the most widely accepted scenario, the
Galactic field population originates in star clusters and associations which 
dissolve in a short timescale \citep{kruijssen2012}. 
Although the initial conditions of such clusters were likely different from those of proto-GCs both in terms of metallicity and gas density, hydrodynamic models of star forming regions suggest that the stellar properties and multiplicity are relatively insensitive to the initial conditions \citep{bate2009a,bate2009b}.

Because of the large initial half-mass radius of the FG and the absence of massive stars in the SG, the escape velocity of the system in the first 30 Myr is smaller than the characteristic kick velocity of SNe II.
So, only a tiny fraction of massive remnants (NS and BH) are retained in our simulations.
To check the effect of a larger retention of massive remnants, in one simulation we switched off the natal kicks of core collapse SNe.

The simulations are run for 12 Gyr allowing to follow the evolution the structural properties of FG and SG and their binary populations.

The whole set of simulations is summarized in Table \ref{tab1_tab}.

\section{Results}
\label{res_sec}

\subsection{General structure}
\label{struc_sec}

\begin{figure*}
 \includegraphics[width=\textwidth]{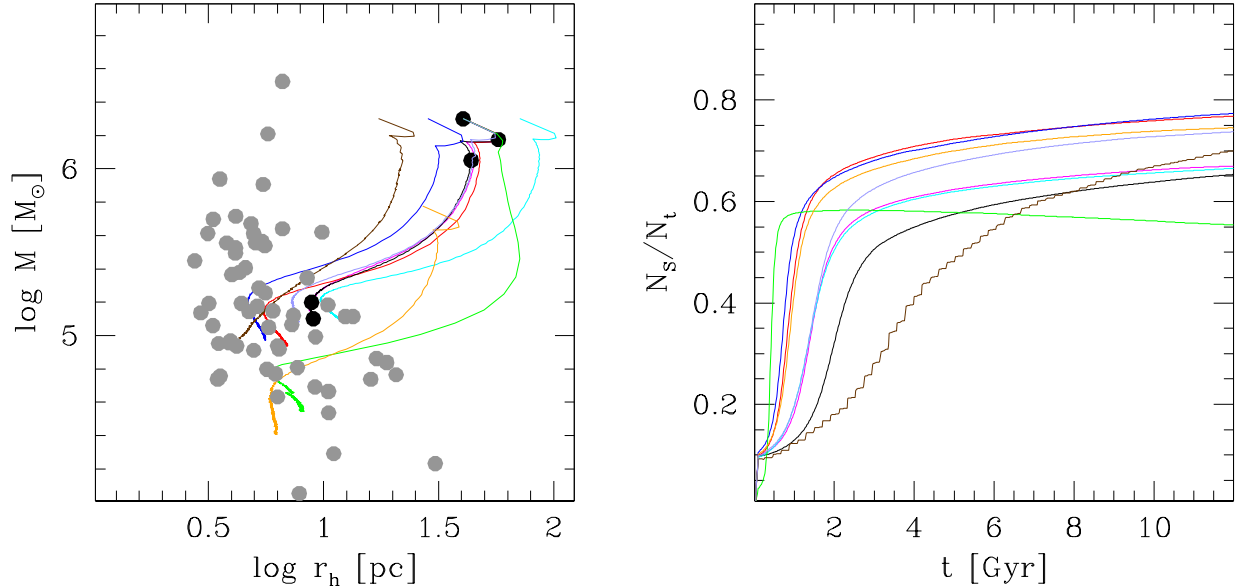}
 \caption{Left panel: Evolution of all the simulations in the $M-r_{h}$ plane. Black points mark the characteristic evolutionary phases described in Sect. \ref{struc_sec} on the R8.5M2fm0.1rs1 simulation. Grey points represents the Galactic GCs with $5<R_{G}/kpc<20$ \citep[from ][]{baumgardt2018}. Right panel: evolution of the fraction of SG stars.
 The colour code is described in the last column of Table \ref{tab1_tab}.}
\label{mr}
\end{figure*}

The simulations presented in this paper are aimed at reproducing the observational evidence in GCs.
As a first sanity check, we show in the left panel of Fig. \ref{mr} the evolution of all the simulations in the $M-r_{h}$ plane.
Qualitatively, all simulations follow a similar path in this diagram: {\it i)} a sudden expansion due to the stellar evolution-driven mass loss ($0<t/Myr<30$), 
{\it ii)} a radial oscillation occurring at the onset of the SG formation which brakes the FG expansion ($30<t/Myr<100$), 
{\it iii)} a strong FG mass-loss corresponding to the shift of the half-mass radius toward the more concentrated SG ($100<t/Myr\lesssim 6000$), {\it iv)} a phase of stable mass-loss during which the cluster maintains a roughly constant size ($6000\lesssim t/Myr<12000$).
It can be seen that, after 12 Gyr of evolution, all the simulations reach a portion of the $M-r_{h}$ plane occupied by low-mass ($M\sim10^{5}~M_{\odot}$) Galactic GCs.
The two simulations starting with a less massive FG (R8.5M0.6fm0.1rs1) or SG (R8.5M2fm0.03rs1) mass finish their evolution with a lower final mass \citep[$M\sim3\times10^{4}~M_{\odot}$; still compatible with those of the least massive GCs; ][]{baumgardt2018}.
While the reason for such a behaviour in the R8.5M0.6fm0.1rs1 simulation is quite obvious (the whole evolution of this simulation is shifted at low masses since its beginning),
the evolutionary path of the R8.5M2fm0.03rs1 simulation is different from that of the other ones.
Indeed, the small braking effect of the low-mass SG leads to a long lasting expansion of the FG, with a corresponding strong mass loss.
Still, the portion of the $M-r_{h}$ plane populated by high-mass GCs (at $log(M/M_{\odot})>5.5$) is not reached by our simulations.
By assuming a rigid shift of the existing simulations, these GCs could originate from massive ($log(M/M_{\odot})>7$) 
progenitors with half-mass radii of $r_{h}\sim 60~pc$. 
Stellar systems of similar masses and sizes have been observed in the periphery of giant elipticals \citep[the Ultra compact dwarfs][]{mieske2002} and in lensed young massive clusters at high redshift \citep{vanzella2019}.

In the right panel of Fig. \ref{mr} the evolution of the number fraction of SG stars is shown for all the performed simulations.
The evolution is characterized by a sudden increase of the SG number fraction due to the early loss of FG.
The fraction of SG reaches an almost constant value of $55-80\%$ after a few Gyr.
This is not surprising since all the performed simulation starts with a large Roche-lobe filling factor $r_{h,F}/r_{J}=0.2$ to reproduce this behaviour.

Summarizing, all the performed simulations finish their evolution with a structure and general properties compatible with those observed in real GCs.
Of course, the set-up of our simulations is quite simplified so that only 
general properties like mass, half-mass radius, and relative fraction of FG/SG are used as constraints, while other 
characteristics (like the correlations among various parameters) would require
a detailed cluster-to-cluster comparison with a fine tuning of the initial conditions.
Nevertheless, the ranges covered by other parameters like the mass function slope ($-0.50<\alpha<-0.88$), the binary fraction ($0.08<f_{b}<0.18$), the half-to-core radii ratio ($0.54<log(r_{h}/r_{c})<1.29$) and the mean velocity dispersion ($1.91<\sigma/km~s^{-1}<3.66$)
are all encompassed by those covered by real GCs in the same mass range \citep{baumgardt2018}.

\subsection{Binaries: general properties}
\label{bingen_sec}

\begin{figure*}
 \includegraphics[width=\textwidth]{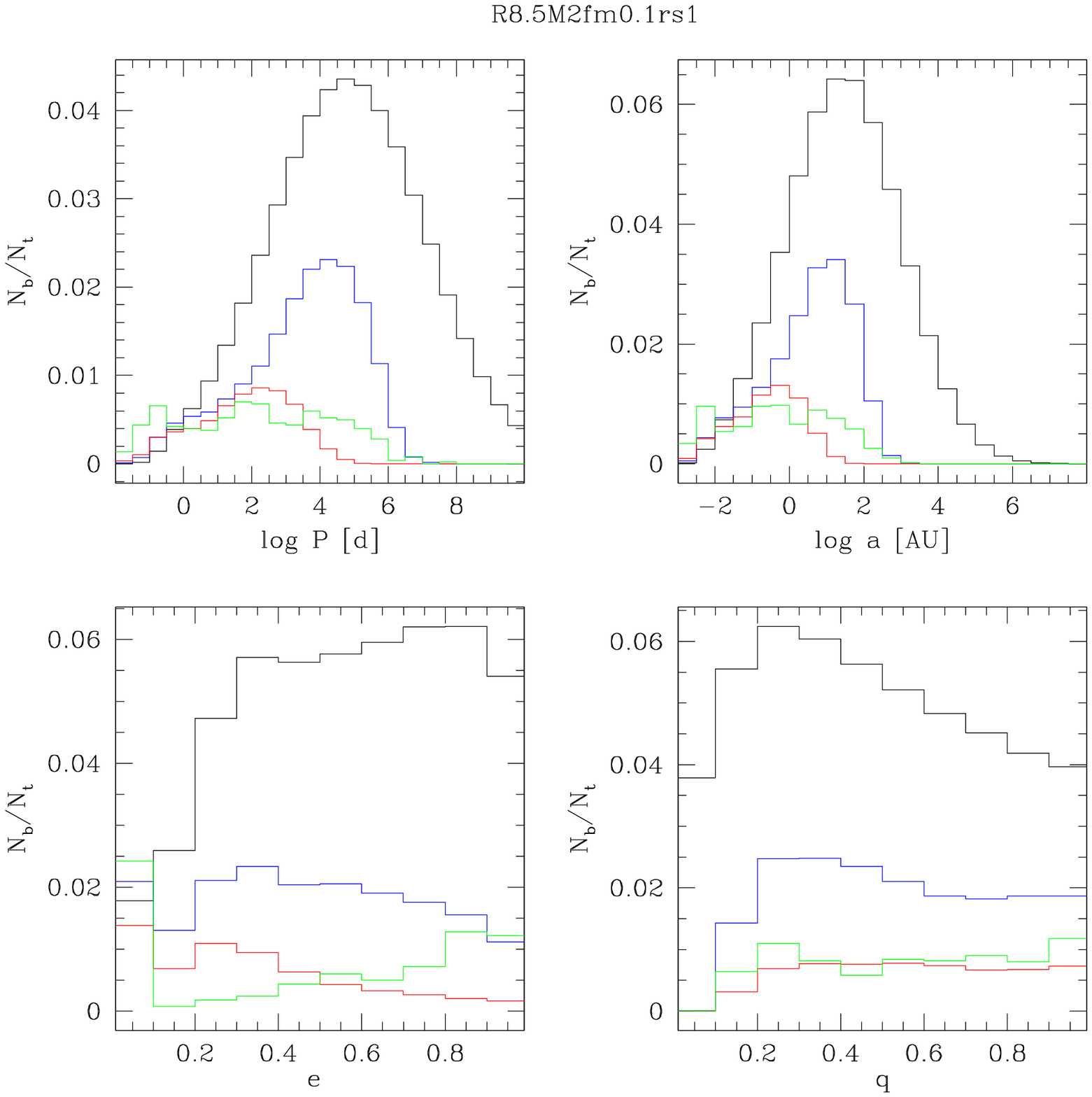}
 \caption{Comparison between the periods (top-left panel), semi-major axis (top-right), eccentricity (bottom-left) and mass-ratio (bottom-right) distributions at the first (black histograms) and the last snapshot of the R8.5M2fm0.1rs1 simulation.
 The blue, red and green lines refer to the FG, SG and mixed binaries, respectively.
 All FG and SG distributions are normalized to the total number of stars of the corresponding populations, while the distributions of mixed binaries are arbitrarily scaled for a better visualization.}
\label{binev}
\end{figure*}

As described in Sect. \ref{setup_sec}, the properties of the binary stars change during the cluster evolution as a result of the joint effect of evaporation and ionization.
Specifically, the global fraction of binaries quickly decreases at the beginning of all simulations as a result of the breaking of soft binaries, reaching values in the range $7\%<f_{b}<19\%$ after 12 Gyr, in agreement with the fractions observed in low-mass GCs by \citet{milone2012}.  
To further illustrate this evolution, we compare in Fig. \ref{binev} the initial and final properties (period, semi-major axis, eccentricity and mass ratio distributions) of the FG and SG binaries in simulation R8.5M2fm0.1rs1.
In this Figure all distributions have been normalized to the total number of objects of the respective population.
The effect of close interactions is apparent from this figure: in both populations the period and semi-major axis distributions are depleted at long periods and large semi-axes.
This range indeed corresponds to those binaries with smaller binding energies ($E_{b}=G m_{1}m_{2}/2 a$), more prone to ionization.
\footnote{The separation is the main parameter driving the binary binding 
energy: as an example, among the binaries contained in the snapshot at 12 Gyr of simulation R8.5M2fm0.1rs1, the semi-major axis varies by a factor 10000 while the product $m_{1}m_{2}$ varies by only a factor 100.}
However, the truncation period and semi-major axis is larger in the FG than in the SG.
This is due to the concentration of SG which populates a region of high density and velocity dispersion where collisions are more frequent and energetic \citep[see also][]{vesperini2011,hong2015,hong2016}.
Other processes of binary creation/destruction like tidal capture, 3-body capture and mergers have a very minor effect on the final binary fractions of FG/SG.
They indeed occur only in extremely dense environments and mainly in massive ($>10~M_{\odot}$) binaries (underrepresented by the adopted bottom-heavy mass function) or in heavy remnants (often expelled through natal kicks and Spitzer instability).

It is interesting to note that also the eccentricity and mass-ratio distributions evolve with time.
The large number of collisions indeed favours exchanges between components, thus enhancing the relative proportion of high mass ratios.
Similarly, at odds with soft binaries, hard binaries release energy to the colliding objects getting harder \citep{heggie1975}.
In this situation tides between the two components becomes important and tend to circularize their orbit.
So, while the initial distribution of eccentricities is close to the thermal one, at the end of the simulation it is peaked at low eccentricities.
Note also that, since the rate of collisions is higher in the SG, the eccentricity distribution of SG binaries is much more peaked than that of FG.

A note of interest comes from mixed binaries (those formed by components of different generations).
These binaries form exclusively from exchanges or tidal capture, whose efficiency is significant only in the central region of the cluster.
Their period and semi-major axis distribution is indeed almost flat with a larger proportion of tight binaries than those observed among FG and SG binaries.
Moreover, their eccentricity distribution is bimodal with a strong peak at circularized binaries (coming from the tidal evolution of tight binaries) superposed to a thermal distribution at large eccentricities (coming from violent perturbation occurring during collisions).
The global fraction of mixed binaries is extremely low (they constitute 1.2\% of all binaries in simulation R8.5M2fm0.1rs1), but it increases in the core (9.4\%), in agreement with what found in the N-body simulations by \citet{hong2019} and observationally by \citet{milone2020}.

\subsection{Binaries: relative fractions in FG/SG}
\label{binfrac_sec}

\begin{figure*}
 \includegraphics[width=\textwidth]{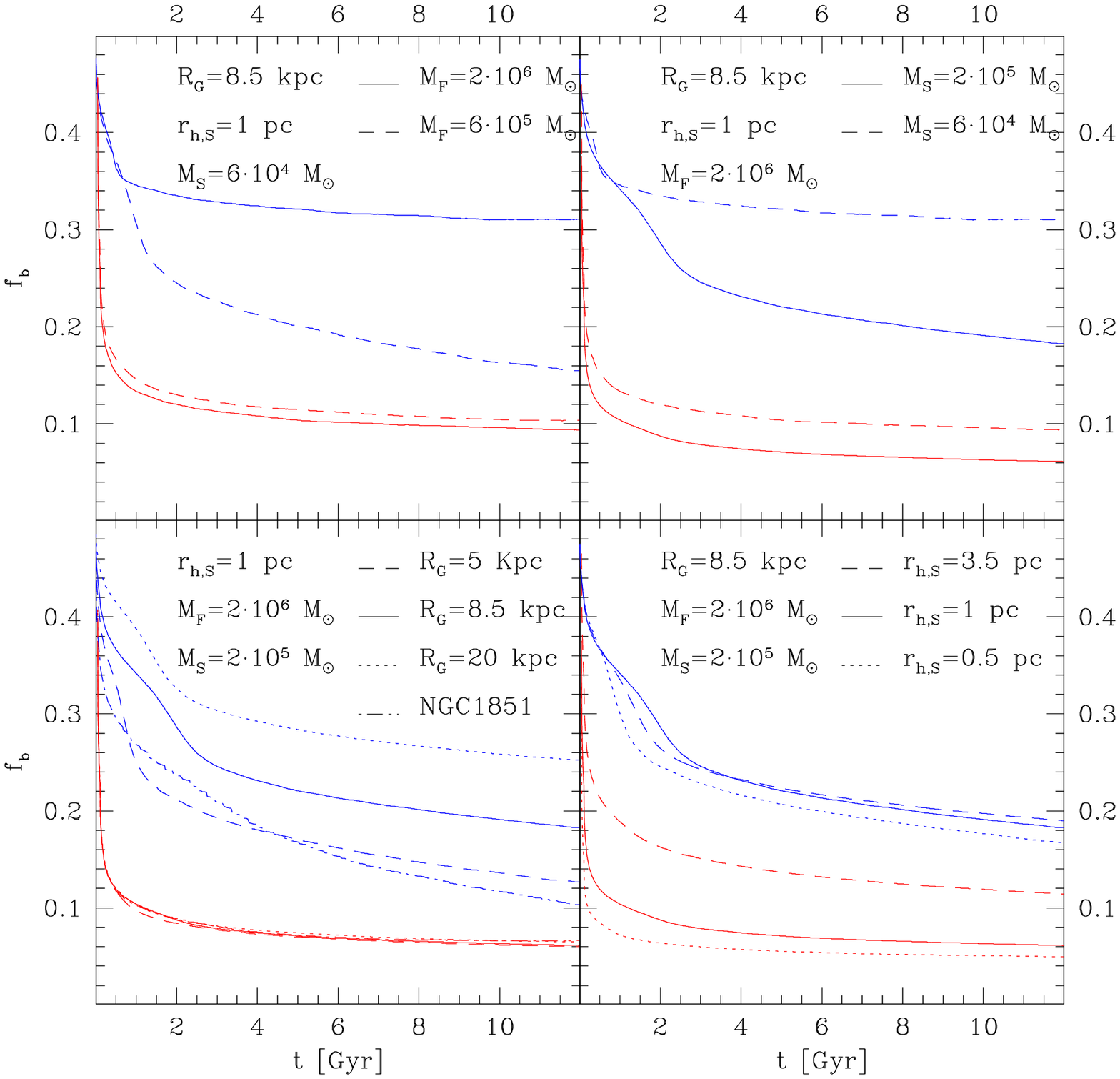}
 \caption{Evolution of FG (blue lines) and SG (red lines) binary fractions in all the simulations.
 Simulations with different initial FG masses (top-left panel), initial SG masses (top-right), orbits (bottom-left) and initial SG half-mass radii (bottom-right) are compared in the various panels.}
\label{fbpar}
\end{figure*}

\begin{figure}
 \includegraphics[width=8.6cm]{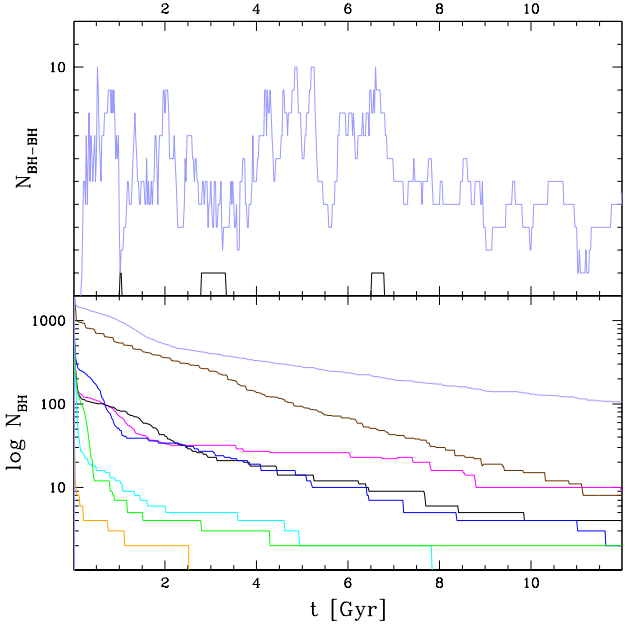}
 \caption{Top panel: evolution of the number of BH-BH binaries in simulations R8.5M2fm0.1rs1 and R8.5M2fm0.1rs1.
 Bottom panel: evolution of the number of BHs in all the simulations.
 The colour code is described in the last column of Table \ref{tab1_tab}}
\label{nbh}
\end{figure}

\begin{figure*}
 \includegraphics[width=\textwidth]{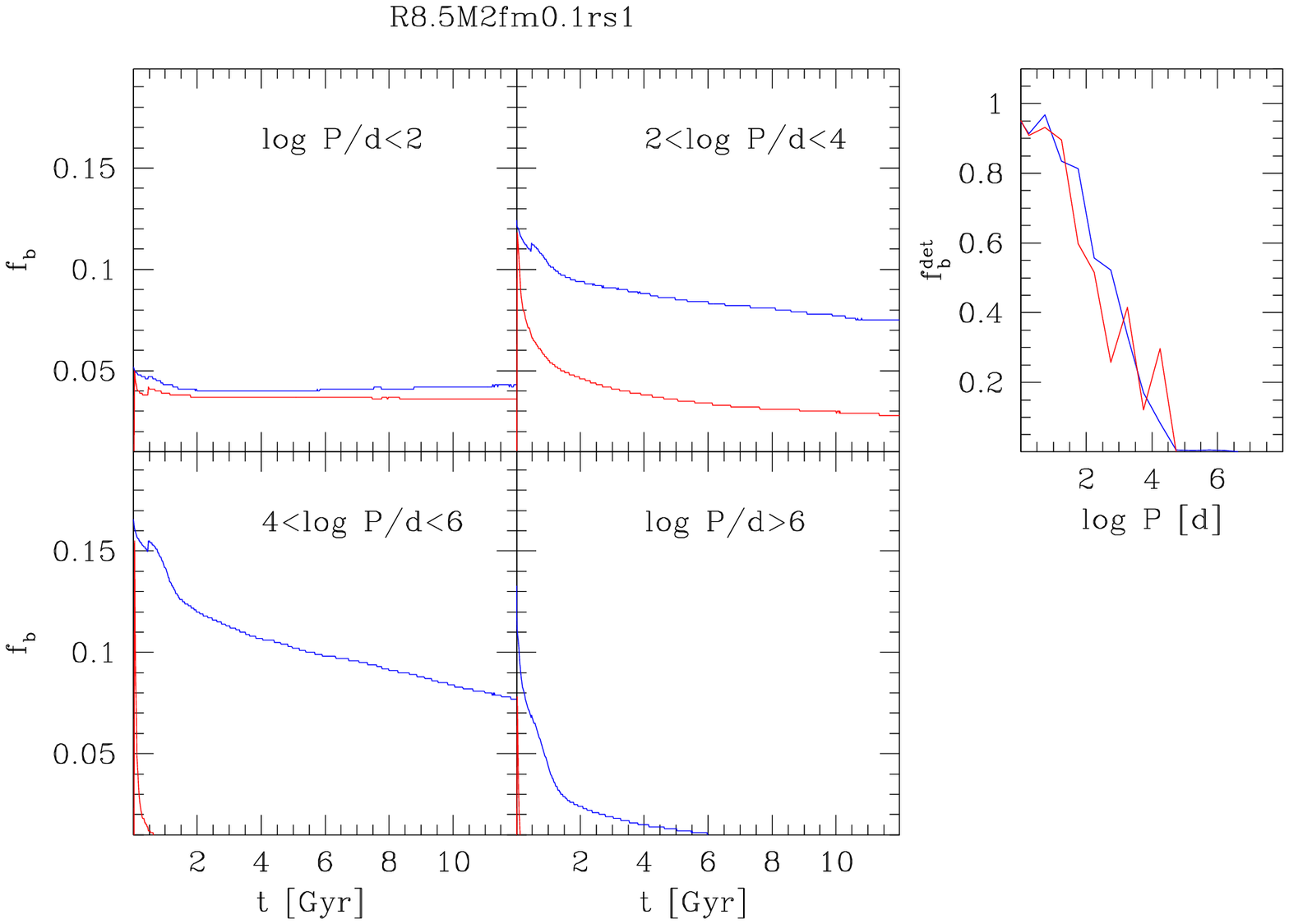}
 \caption{Evolution of the fraction of FG (blue lines) and SG (red lines) binaries in the simulation R8.5M2fm0.1rs1. Left panels refer to binaries in three different period intervals. The right panel show the detection efficiency of the \citet{lucatello2015} analysis.}
\label{fbp}
\end{figure*}

In this Section we analyse the effect of various initial conditions on the binary fractions of FG and SG.
In Fig. \ref{fbpar} the evolution of the global fraction of FG and SG binaries in simulation with different FG and SG masses and half-mass radii are compared.

From these comparisons it appears that:
\begin{enumerate}
\item{The fraction of SG binaries ($5\%<f_{b,S}<12\%$) is systematically smaller than that of FG ($10\%<f_{b,F}<32\%$);}
\item{The fraction of binaries drops in the first 0.5 Gyr for the SG and slightly later in the FG, reaching in both cases an equilibrium value which is maintained till the end of the simulation;}
\item{The final fraction of SG binaries decreases as the initial half-mass radius of SG decreases, and increases in the simulations with a low mass FG or SG;}
\item{The final fraction of FG binaries decreases as the Galactocentric distance decreases, in the simulation moving on an eccentric orbit, in that with a low mass FG and increases in that with a low mass SG.}
\end{enumerate}

A Pearson permutation test exploring mono- and bivariate correlations of the SG binary fraction $f_{b,S}$ with the parameters $R_{G},~r_{h,S},~M_{F}$ and $M_{S}$ indicates a unique significant ($P\sim 98.2\%$) correlation between $f_{b,S}$ and $r_{h,S}$ and two marginally significant ($87\%<P<95\%$) bivariate correlations between $f_{b,S}$, $r_{h,S}$ and either $M_{F}$ or $M_{S}$.
The same analysis applied to the FG return a less clear result with no significant correlations. 

The observed trend of the final FG and SG global binary fractions with the various parameters can be interpreted considering the environmental conditions (density, velocity dispersion) of the regions where these two populations spend most of their evolution.
The SG forms in the centre and maintains a compact structure during all its evolution, following a dynamical evolution decoupled from the FG.
Being almost insensitive to the tidal field and to the structural evolution of the FG,
the only parameter affecting the binary fraction is the efficiency of the ionization process.
As described in Sect. \ref{setup_sec}, this efficiency is linked to the fraction of hard binaries i.e. those with a binding energy larger than the local kinetic energy of colliding stars.
This last quantity can be quantified by the squared r.m.s. velocity of stars in the central region where this population resides during all its evolution.
The dependence of such a velocity from the general cluster parameters (mass and half-mass radius) can be derived from the virial theorem so that 
\begin{equation}
\sigma^{2}\propto M/r_{h,S}
\label{sig_eq}
\end{equation}
The larger is this ratio the smaller is the fraction of binaries able to survive to collisions. 
Note that the maximum destruction efficiency of SG binaries occurs at the beginning of the simulation when the SG has its smallest half-mass radius and the cluster is still extremely massive.
In this initial phase, almost all the soft binaries are quickly destroyed, while the harder ones survive until the end of the simulation.

The same mechanism can explain also the variation of the FG binary fraction on the various simulations.
However, at odds with SG, the mass and half-mass radius of the FG vary by about one order of magnitude during the first Gyrs of evolution.
So, the moment of maximum efficiency of the binary ionization process occurs at a later epoch when, as a result of the early mass-loss, the FG half-mass radius shrinks.
This effect is further enhanced by mass segregation of FG binaries which move toward the central region where the dense SG environment contributes to an efficient disruption of binaries.
The fraction of binaries is therefore more sensitive to those parameters which determine the FG structural evolution.
In this regard, a strong tidal field (in simulations moving at small distances from the Galactic centre and/or when moving on eccentric orbits) imposes a smaller initial half-mass radius to guarantee the cluster survival, with a consequent high efficiency of binary ionization.
Similarly, when the FG has a low mass (in simulation R8.5M0.6fm0.1r1), the initial FG half-mass radius must be large enough to ensure a significant loss of FG stars, and the corresponding efficiency of binary ionization is low.
On the other hand, when a small SG forms (in simulation R8.5M2fm0.03r1), the braking effect on the FG is small and the cluster spends most of its evolution with a large half-mass radius, with a corresponding low efficiency of the binary ionization process.

It is interesting to study the effect of BHs for the evolution of the binary fractions of FG and SG.
In the bottom panel of Fig. \ref{nbh} we show the evolution of the number of BH in our simulations.
As expected, the number of BH has a sudden decrease at the beginning of all simulations resulting from the prompt ejection caused by natal kicks.
A continuous loss of BH occurs during the subsequent evolution as a result of the interaction between BH and binaries in the cluster centre, so that less than 10 BH are present at the end of all simulations.
This prediction could conflict with recent indications of a significant population of heavy remnants in the core of two nearby GCs \citep{vitral2022}.
A different situation emerges in simulation R8.5M2fm0.1rs1nk which is run without SNeII natal kicks: in this simulation a significant number of BH is retained from the beinning of the simulation. 
They quickly sink into the cluster central region as a result of mass segregation and interact with other single and binary stars, mainly SG stars residing in the cluster centre.
A fraction of them form binaries through three-body capture and later expell the low-mass companions in repeated exchanges.
The result of this process is the formation of a sub-system of BH-BH binaries acting as a reservoir of energy which is released to the interacting stars.
This is illustrated in the bottom panel of Fig. \ref{nbh} where the number of BH-BH binaries of simulations R8.5M2fm0.1rs1 and R8.5M2fm0.1rs1nk are compared.
It is apparent that while in simulation R8.5M2fm0.1rs1 BH-BH binaries are rare, simulation R8.5M2fm0.1rs1nk hosts a roughly constant number of $4-10$ BH-BH binaries during the entire evolution.
The interaction of BH-BH binaries with SG stars lead to an increase of the half-mass radius of the SG. 
In this configuration, the ionization rate of SG stars decreases and a slightly larger fraction of binaries is maintained with respect to the simulation where BH-BH binaries are absent.
This process is not significant for FG binaries which are distributed over a large area, outside the range of action of BH-BH binaries.

Note that the difference between the binary fractions in the two generations decreases at short periods and small semi-major axes.
In fact, in this range binaries of both populations are generally hard and their fractions are less sensitive to the environment.
To better visualize this effect, we show in Fig. \ref{fbp} the evolution of the binary fractions of FG and SG in four period ranges.
It is apparent that, while at intermediate periods ($2<log P/d<6$) the two generations show a strikingly different destruction efficiency, outside this range binaries are either easily disrupted (at $log P/d>6$) or preserved (at $log P/d<2$) in both populations.

To compare the binary fractions in our set of simulations with the observational results of \citet{lucatello2015} it is necessary to account for the observational bias of this last work.
In fact, because of the limited duration of the radial velocity monitoring of this study, the efficiency of binary detection varies as a function of the binary orbital period.
In particular, long-period binaries are characterized by small amplitude variations possibly hidden in the observational uncertainty, and the radial velocity monitoring could have sampled only a small fraction of the orbital phase.
Observational analyses based on spectroscopic time-series are therefore progressively less sensitive at increasing periods \citep[see Fig. 4 of][]{lucatello2015}. 
To account for this effect, we select from the last snapshot of each simulation only those binaries with an unevolved primary component with $m_{1}>0.8~M_{\odot}$ \citep[corresponding to the Red Giant mass of the GCs analyzed by][]{lucatello2015}.
For each of these binaries, we assumed a random phase of maximum, inclination angle and longitude of the periastron and calculated the radial velocity of the primary component with respect to the binary center of mass at the orbital phase corresponding to the cadence of the spectroscopic observations of \citet{lucatello2015}.
A gaussian shift with standard deviation equal to the radial velocity uncertainty has been added to velocities to mimic the effect of observational uncertainties.
Following the above procedure, a mock observation with the same observational bias of the \citet{lucatello2015} study has been simulated.
The detection of binaries has been then performed adopting the same procedure described in \citet{lucatello2015}.
A correction has been applied to account for the biased radial distribution of spectroscopic targets \citep[see Fig. 1 of ][]{lucatello2015}.
For this purpose, to each particle of the simulation, a weight equal to the ratio between the number of spectroscopic targets and that of the expected stars within $\Delta log~r/r_{h}<0.25$ from the particle location has been assigned.
The fractions of observable FG and SG binary fractions ($f_{b,F}^{obs}$ and $f_{b,S}^{obs}$, respectively) have been determined as the ratio between the sum of the weights of detected binaries and of all systems in the corresponding population, and are listed in Table \ref{tab1_tab}.

By restricting the analysis to observable binaries the range covered by the various simulations reduces significantly spanning intervals of $4\%<f_{b,F}^{obs}<10$\% and $1.5\%<f_{b,S}^{obs}<7$\%.
The same dependences outlined for global fractions holds also for this subsample of binaries.
In particular, FG binaries are more numerous than SG ones with a fraction that increase as the tidal field becomes weaker.
The dispersion of FG observable binary fractions is however much smaller than that of the global fraction ($\sigma f_{b,F}^{obs}/f_{b,F}^{obs}=0.28$; $\sigma f_{b,F}/f_{b,F}=0.36$).
This is due to the fact that observable binaries are preferentially short-period (mainly hard) binaries, whose fraction is more stable (see Fig. \ref{fbp}).  
Taken at face value, the fraction of FG observable binaries in almost all the simulations moving on circular orbits is slightly higher than that observed by \citet{lucatello2015} ($f_{b,F}=4.9\pm1.3\%$).
However, most GCs follow eccentric orbits experiencing a tidal field stronger than that felt by our simulations moving on circular orbits without experiencing any disk/bulge-shock.
Indeed, the FG binary fraction significantly reduces ($f_{b,F}^{obs}=3.8\%$) in simulation 1851M2fm0.1r1 which moves within a realistic potential on an eccentric orbit, dropping below the value measured by \citet{lucatello2015}.

Regarding the SG, the best match with the value measured by \citet{lucatello2015} ($f_{b,S}=1.2\pm0.4\%$) is obtained by the simulations R8.5M2fm0.1rs0.5 ($f_{b,S}^{obs}=2.3\%$) and R8.5M2rhrj0.2fm0.03rs1ms ($f_{b,S}^{obs}=1.7\%$).
Similarly to the behaviour observed with the global fractions, there is a clear trend of increasing SG binary fraction with SG initial half-mass radius, up to $f_{b,S}^{obs}=5.3\%$ at $r_{h,S}=3.5~pc$ (R8.5M2fm0.1rs3.5).
Note that a similar binary fraction ($f_{b,S}^{obs}=5.1\%$) is measured in simulation R8.5M0.6fm0.1rs1 which, in spite of its small initial SG half-mass radius, is characterized by a small initial FG mass.
So, we conclude that simulations sharing the same $M_{F}/r_{h,S}$ ratio (and therefore similar velocity dispersion within the SG half-mass radius, see eq. \ref{sig_eq}) also have similar fractions of SG binaries.
Indeed, a two-dimensional fit in the plane of the correlated parameters ($f_{b,S}^{obs},~r_{h,S},~M_{F}$) gives
\begin{equation}
log~f_{b_S}^{obs}=0.47~log~\left(\frac{r_{h,S}}{pc}\right)-0.52~log~\left(\frac{M_{F}}{10^{6}M_{\odot}}\right)-1.41
\label{fbs_eq}
\end{equation}
indicating that $f_{b_S}^{obs}\propto \sqrt{r_{h,S}/M_{F}}$.
Unfortunately, none of the performed simulations reaches an observed SG binary fraction as low as that measured by \citet{lucatello2015}.
However, following the above considerations, it is possible to invert eq. \ref{fbs_eq} to extrapolate the initial half-mass radius of SG (at the epoch of its formation $t_{S}$) as a function of the initial cluster mass ($M(t=0)\equiv M_{F}$) in order to match the observational SG binary fraction as
\begin{equation}
r_{h,S}(t_{S})=0.1_{-0.03}^{+0.04} \left(\frac{M(t=0)}{10^{6}~M_{\odot}}\right)~pc
\label{constr_eq}
\end{equation}
where the associated uncertainty is calculated from the propagation of the observational uncertainty and the intrinsic scatter of eq. \ref{fbs_eq}.
Other sources of systematic uncertainties linked to the uncertainties in the adopted distribution of binary binding energies, to the structural differences between simulations and real GCs and/or to the systematics possibly affecting the observational estimate could be also present.
So, a more conservative $3\sigma$ upper limit to the SG half-mass radius could be given as
\begin{equation}
r_{h,S}(t_{S})<0.22 \left(\frac{M(t=0)}{10^{6}~M_{\odot}}\right)~pc
\label{constr2_eq}
\end{equation}
 


\section{Conclusions}
\label{concl_sec}

\begin{figure}
 \includegraphics[width=8.6cm]{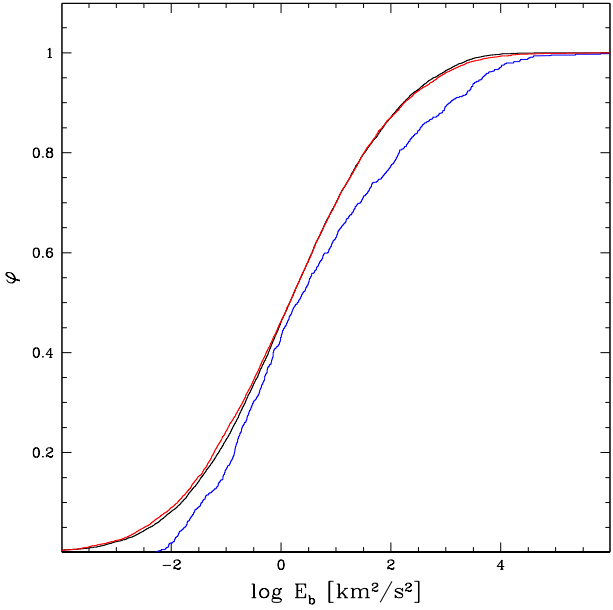}
 \caption{Cumulative distribution of initial binding energies of binaries according to the prescriptions of \citet{duquennoy1991} (black line), \citet{raghavan2010} (red line) and \citet{moe2017} (blue line).}
\label{eb}
\end{figure}

In this paper we performed a set of Monte Carlo simulations of star clusters composed by two stellar populations, accounting for the main processes affecting their dynamical evolution (stellar evolution, tides, two-body relaxation, three- and four-body interactions) and exploring a range of initial sizes and masses of the two populations.
These simulations host an initial fraction of binaries compatible with that measured in the Galactic field \citep[$f_{b}=50\%$; ][]{duquennoy1991,moe2019} and quickly destroy a fraction of them with an efficiency which depends on their initial conditions.
We observe that wide ($a>10~AU$) binaries are quickly destroyed in the central portion of the cluster where the SG resides, creating a difference in the binary fractions of FG and SG which is maintained during the subsequent evolution.
Among the various simulations, we found that the fraction of SG binaries depends uniquely on the ratio between the initial cluster mass and the half-mass radius of the original nucleus of SG stars.
This is a consequence of the dependence of the r.m.s. velocity of stars on the enclosed mass and size of the region containing the SG, which determines the boundary between soft and hard binaries. 
The efficiency of binary ionization decreases with time as a result of the sudden mass loss experienced by the cluster and by the expansion of the SG during its dynamical mixing process, so that the final binary fraction in the SG observed today is strongly sensitive to the initial size of the SG.
{\it This indicates that the SG binary fraction represents an excellent tool to constrain the initial concentration of SG at its birth.}
In particular, through a comparison with the observational work of \citet{lucatello2015} after taking into account for observational effects, we provide an estimate of such an initial size, which depends uniquely on the initial cluster mass (see eq.s \ref{constr_eq} and \ref{constr2_eq}).
Considering that GCs have present-day masses 2$\div$4 times larger than those reached by our simulations after 12 Gyr of evolution \citep{baumgardt2018}, and assuming a constant initial-final mass ratio, we argue that SG must have formed with typical half-mass radii smaller than 0.5-1 pc.
In spite of the associated uncertainties, this is the first time that a direct link between initial conditions and a present-day parameter is identified.
Indeed, all the other dynamical quantities are strongly affected by the effect of two-body relaxation which erases primordial differences between FG and SG.

The fraction of FG binaries shows instead a more complex dependence on the cluster orbital parameters.
Indeed, at odds with SG, FG is more extended and exposed to the tidal field during the entire cluster evolution.
So, the r.m.s. velocity of colliding stars ($\sigma\propto \sqrt{M/r_{h}}$), determining the soft-hard boundary, follows a slow evolution with
the contraction of FG partly compensates its strong mass loss.
So, while very soft (corresponding to wide binaries with separations of $a>100~AU$ for a typical combination of components' masses) binaries are immediately destroyed, intermediate ones (with $10<a/AU<100$) take a longer timescale to be ionized.
The final value of the FG binary fraction therefore depends, beside structural parameters, on the strength of the tidal field which determines the variation of cluster mass and size. 
On the other hand, when the analysis is restricted to the sample of tight binaries ($a<10~AU$), a more stable fraction of FG binaries is measured among the various simulations.
These binaries are indeed more resistant to ionization and their fraction is therefore less dependent on external factors.

It is important to point out that our results and constraints are based on a number of assumptions.
First, we set up our simulations according to the scenario envisaged by \citet{dercole2008}.
It cannot be taken for granted that the same result holds also in a different framework.
On the other hand, as far as the SG forms in a timescale shorter than the duration of the binary ionization process ($<100-500~Myr$), the actual nature of the polluter should not affect these results.
So, the conclusions drawn here should remain valid for all those models predicting the SG formed from an early recycle of polluted gas in the central region of the FG \citep[e.g.][]{dercole2008,krause2013}.
Alternative scenarios could also lead to a present-day structure similar to that 
observed among GCs e.g. starting from compact configurations of FG and SG \citep[e.g.][]{bastian2013,gieles2018}.
While these scenarios needs to be tested with specifically designed simulations, some of them have been found to lead to a few inconsistencies with observations \citep{sollima2021}.

Second, the quantitative constraint on the initial size of SG depends on the adopted characteristics of binaries (period, semi-major axis, eccentricity, mass-ratio distribution).
All these parameters contribute to the binary binding energy, so that different choices lead to populations of binaries with a different resistance to ionization. 
In this paper, we adopted the classical prescriptions of \citet{duquennoy1991}, but alternative distributions have been proposed in recent years \citep{raghavan2010,sana2012,moe2017}. 
It is worth mentioning that all these works predict very similar binding energy distributions differing by $<10\%$ across the entire range covered by the binaries in our simulations (see Fig. \ref{eb}).
While this represents an unavoidable source of uncertainty, a realistic model able to predict the characteristics of binaries in complex environments like those of GC precursors is still missing.
As already discussed in Sect. \ref{setup_sec}, hydrodynamical simulations suggest only a weak (if any) dependence of the binary properties on environmental conditions \citep{bate2009a}, which should be therefore determined by dynamical evolution.
On the other hand, the extremely different density expected in proto-GCs is expected to affect the relative efficiency of accretion and fragmentation in proto-stellar accretion disks as well as their sizes and lifetimes, which affect the eccentricity and mass-ratio distributions of binaries \citep{tokovinin2020}.
We will address the dependence of the binary properties on the density and metallicity of the environment in a future work. 

Third, we adopted a simplified tidal field in our simulations to mimic clusters subject to various degrees of tidal stress.
In real cases, the effect of bulge- and disk-shocking could be significant in GCs following orbits with a large eccentricity or confined close to the Galactic disk.
This is particularly important for FG binaries whose fraction is more sensitive to the tidal field or, more specifically, to the effect of the tidal field on the structure of the FG system..
In this regard, it is interesting to note that simulation 1851M2fm0.1rs0.1, run following an eccentric orbit in a realistic Galactic potential, provides a FG binary fraction in better agreement with the observed measure \citep{lucatello2015} with respect to simulations moving on circular orbits in a logarithmic potential.

The result presented here agrees with the qualitative prediction of \citet{gratton2019} which was based on a simplified toy model.
The constraint to the half-mass radius of the SG determined here implies a very high half-mass density for the SG ($\rho_{h,S}> 10^{5} M_{\odot}~pc^{-3}$) decreasing by about one order of magnitude in the first 100 Myr. 
Similar densities are predicted by the hydrodynamic simulations of \citet{calura2019} (see their Fig. 3). 
These authors performed their simulations in the multiple-generations scenario and predict a SG forming with $\rho_{S}\sim10^{6} M_{\odot}pc^{-3}$.
Observationally, the typical density of the site of star formation are much lower than this value \citep{hopkins2010}.
The known embedded star clusters in the Milky Way contains $\sim 10^{2\div 4}M_{\odot}$ in regions with effective radii of $\sim 1$ pc \citep{lada2003}.
High-density peaks can be found in the local Universe around Young Massive Clusters \citep[with $\rho\sim10^{3}~M_{\odot}~pc^{-3}$; ][]{portegieszwart2010}, still orders of magnitude smaller than that estimated here. 
However, all these young clusters are significantly less massive and located in environments characterized by higher metallicity than GC progenitors, so that star formation could have proceeded in a different way.
Densities of the order of $10^{6}M_{\odot}pc^{-3}$ have been found in the nuclear star clusters of the Milky Way and other late-type galaxies \citep{schodel2014,georgiev2016}.
On the other hand, extremely dense star-forming regions have been recently observed in lensed fields at high redshift by \citet{vanzella2019} which could be associated to GC precursors.

Unfortunately, the present study suffers from the weak constraint provided by the only available observational study performed on several GCs \citep{lucatello2015}.
This pioneering work used a dataset of spectra collected over $\sim$3 years in different observing campaigns aimed at the determination of chemical abundances, without an optimal cadence and surveyed only a few tens of Red Giants per cluster, thus requiring to stack the samples of all GCs together.
In this situation, the combined effect of individual peculiarities, different radial sampling, structural differences, etc. adds further uncertainty to the estimated binary fractions. 
Future studies specifically devoted to the determination of binary fractions with a better statistics and a higher sensitivity to longer periods will help to improve the constraint provided in this paper.

\section*{Acknowledgements}

We warmly thank Enrico Vesperini and Francesco Calura for the useful discussions and Holger Baumgardt for providing his N-body simulations.
We also thank Mirek Giersz, the referee of our paper, for his helpful comments and suggestions that improved our paper.

\section*{Data Availability}


The data underlying this article will be shared on reasonable request to the corresponding author.

\begin{landscape}
\begin{table}
  \caption{Initial and final properties of the performed simulations.}
  \begin{tabular}{@{}lcccccc@{\hskip 1cm}cccccccl@{}}
 \hline
                  & \multicolumn{5}{c}{Initial} & \multicolumn{7}{c}{Final} &\\
 name             & $R_{G}$ 	 & $r_{h,F}$ & $r_{h,S}$ & $M_{F}$	        & $M_{S}/M_{tot}$ & natal &M                   & $r_{h}$ & $N_{S}/N_{tot}$ & $f_{b,F}$ & $f_{b,S}$ & $f_{b,F}^{obs}$ & $f_{b,S}^{obs}$ & colour\\				      
                  &	    	 &	     &	         &		        & 		  & kicks &                    &         &                 &	       &	   &		     &  	       &\\				     
                  & kpc	    	 & pc	     & pc	 & $10^{5}~M_{\odot}$   & 		  &       & $10^{5}~M_{\odot}$ & pc      &                 &	       &	   &		     &  	       &\\				     
\hline
 R8.5M2fm0.1rs1   & 8.5     	 & 37.18     & 1	 & 20		     	& 0.1		  & yes   & 1.47	       &  9.95   & 0.65 	   & 0.183     & 0.061     & 0.078	     & 0.025	       & black\\  
 R8.5M2fm0.1rs0.5 & 8.5     	 & 37.18     & 0.5	 & 20		     	& 0.1		  & yes   & 0.97	       &  7.41   & 0.77 	   & 0.167     & 0.050     & 0.064	     & 0.023	       & red\\    
 R8.5M2fm0.1rs3.5 & 8.5     	 & 37.18     & 3.5	 & 20		     	& 0.1		  & yes   & 1.71	       &  9.34   & 0.67 	   & 0.190     & 0.114     & 0.082	     & 0.053	       & magenta\\
 R5M2fm0.1rs1     & 5       	 & 26.10     & 1	 & 20		     	& 0.1		  & yes   & 1.05	       &  5.83   & 0.77 	   & 0.126     & 0.060     & 0.071	     & 0.025	       & blue\\   
 R20M2fm0.1rs1    & 20      	 & 65.78     & 1	 & 20		     	& 0.1		  & yes   & 1.49	       & 14.11   & 0.67 	   & 0.252     & 0.065     & 0.095	     & 0.040	       & cyan\\   
 R8.5M0.6m0.1rs1  & 8.5     	 & 24.89     & 1	 &  6		     	& 0.1		  & yes   & 0.30	       &  6.43   & 0.75 	   & 0.155     & 0.104     & 0.057	     & 0.051	       & orange\\ 
 R8.5M2fm0.03rs1  & 8.5     	 & 37.18     & 1	 & 20		     	& 0.03  	  & yes   & 0.45	       &  9.27   & 0.55 	   & 0.311     & 0.094     & 0.044	     & 0.017	       & green\\  
 R8.5M2fm0.1rs1nk & 8.5          & 37.18     & 1	 & 20		    	& 0.1		  & no    & 1.40               &  8.24	 & 0.74		   & 0.167     & 0.079	   & 0.083	     & 0.047 	       & indigo\\
 1851M2fm0.1rs1   & NGC1851-like & 16        & 1	 & 20		    	& 0.1		  & yes   & 1.03	       &  4.37   & 0.70 	   & 0.103     & 0.066     & 0.038	     & 0.027	       & brown\\
\hline
\end{tabular}
 \label{tab1_tab}
\end{table}
\end{landscape}



\bibliographystyle{mnras}



\appendix

\section{Comparison with N-body simulations}

\begin{figure*}
 \includegraphics[width=\textwidth]{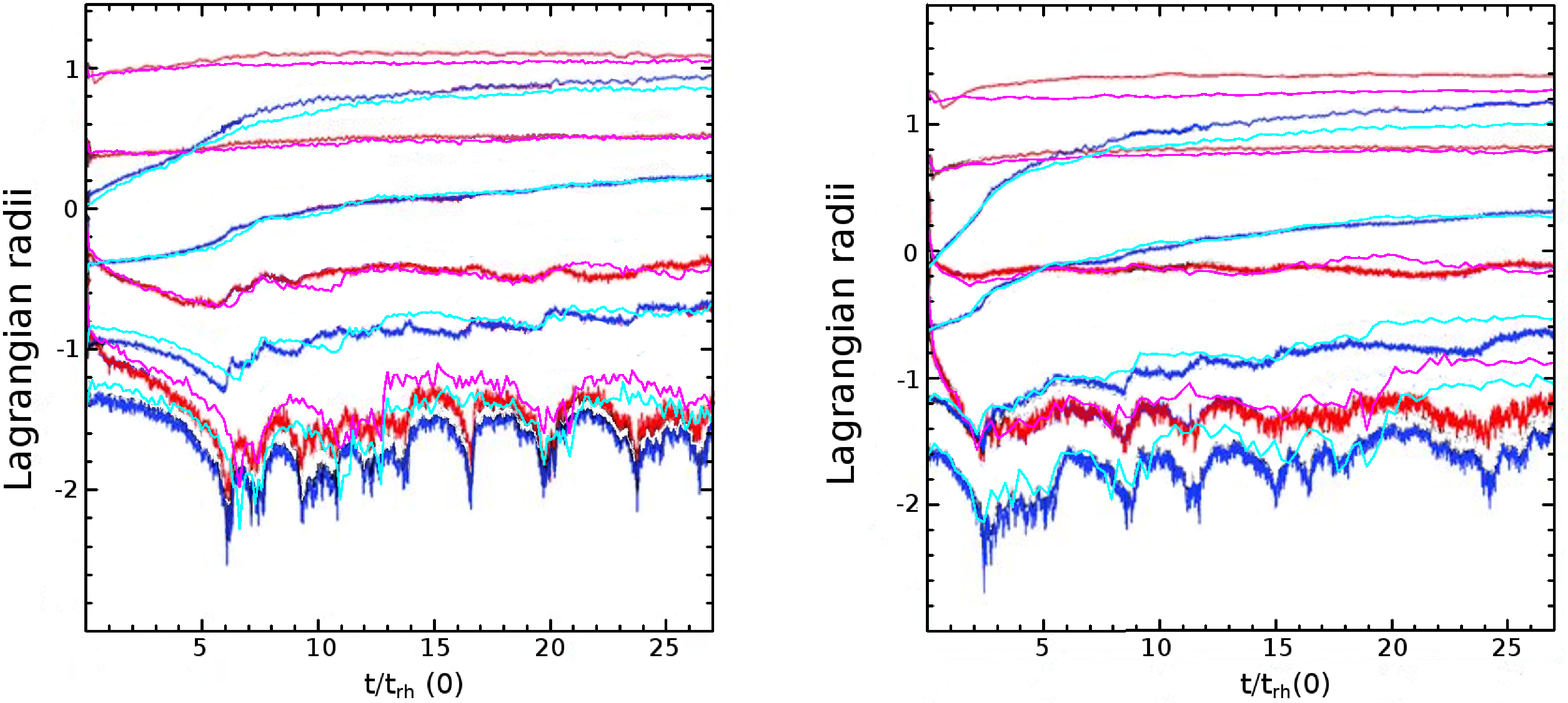}
 \caption{Evolution of the Lagrangian radii containing 1\%, 10\%, 50\% and 90\% of the enclosed mass in the N-body simulations r10 (left panel) and r25 (right panel) of \citet{vesperini2013} and the corresponding Monte Carlo simulations.
 In both panels, the Lagrangian radii of FG and SG are marked by red and blue lines, respectively, for N-body simulations and by magenta and cyan lines for Monte Carlo simulations.}
\label{mcnb_vesp}
\end{figure*}

\begin{figure}
 \includegraphics[width=8.6cm]{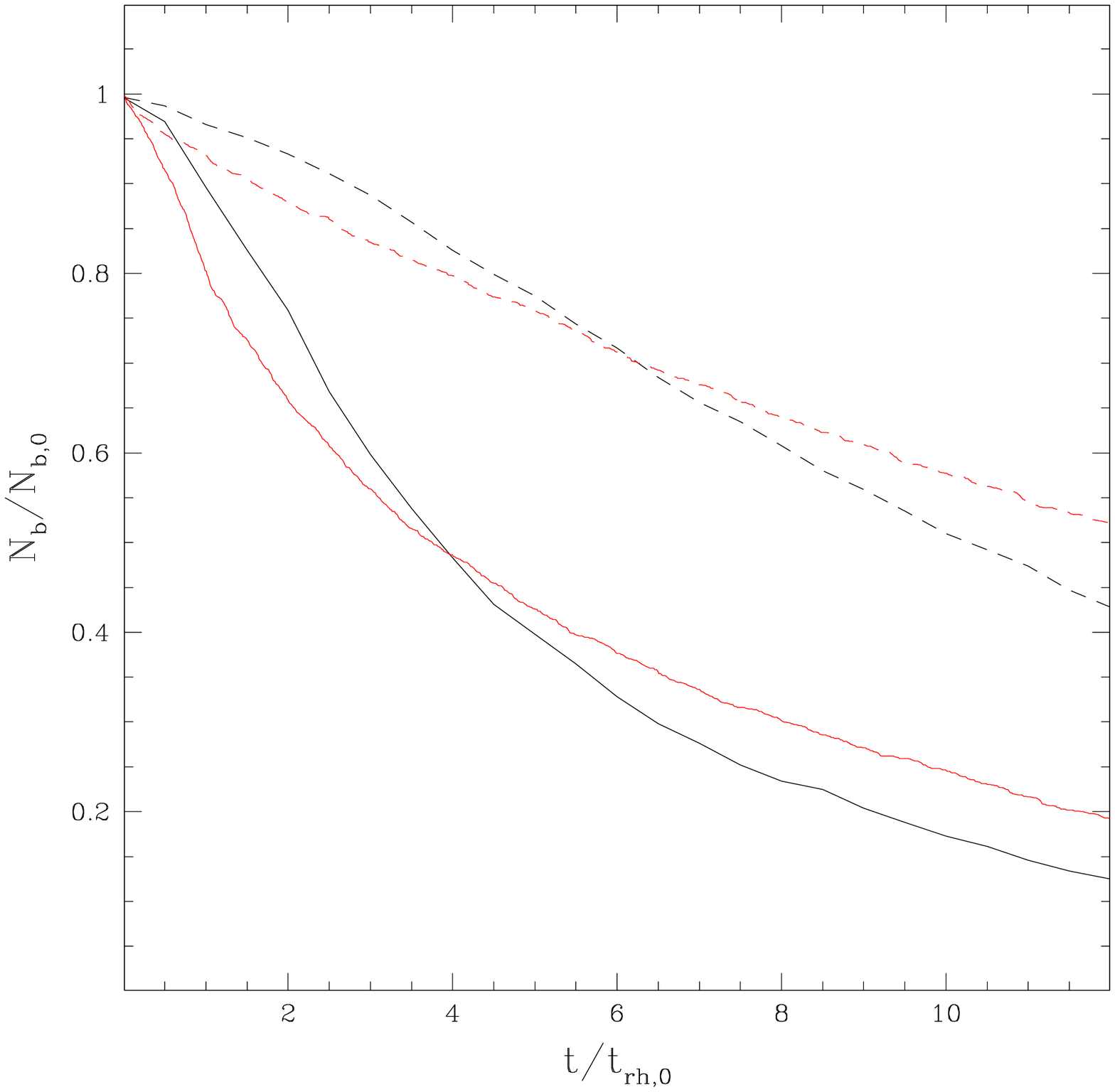}
 \caption{Evolution of the number of binaries in the two N-body simulations of the \citet{hong2015} sample (black lines) and the corresponding Monte Carlo simulations (red lines).
 All numbers are normalized to their initial values. Simulations with hardness x=3 and 20 are marked by solid and dashed lines, respectively.}
\label{mcnb_hong}
\end{figure}

\begin{figure}
 \includegraphics[width=8.6cm]{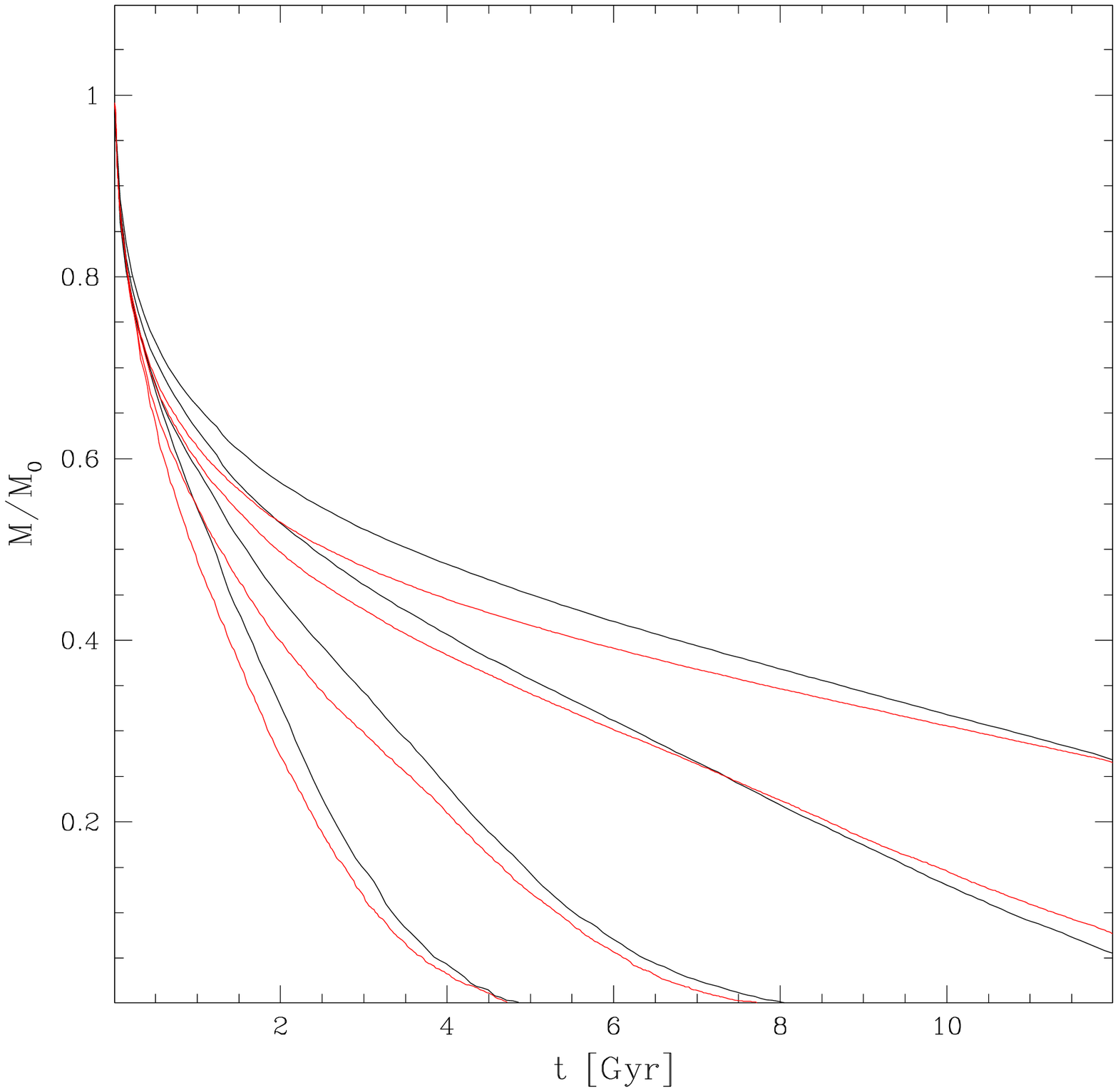}
 \caption{Evolution of the enclosed mass in the four N-body simulations of the \citet{baumgardt2003} sample (black lines) and the corresponding Monte Carlo simulations (red lines).
 All masses are normalized to their initial values. Simulations with N=10000, 20000, 50000 and 100000 particles end their evolution at increasing times.}
\label{mcnb_m}
\end{figure}

\begin{figure*}
 \includegraphics[width=\textwidth]{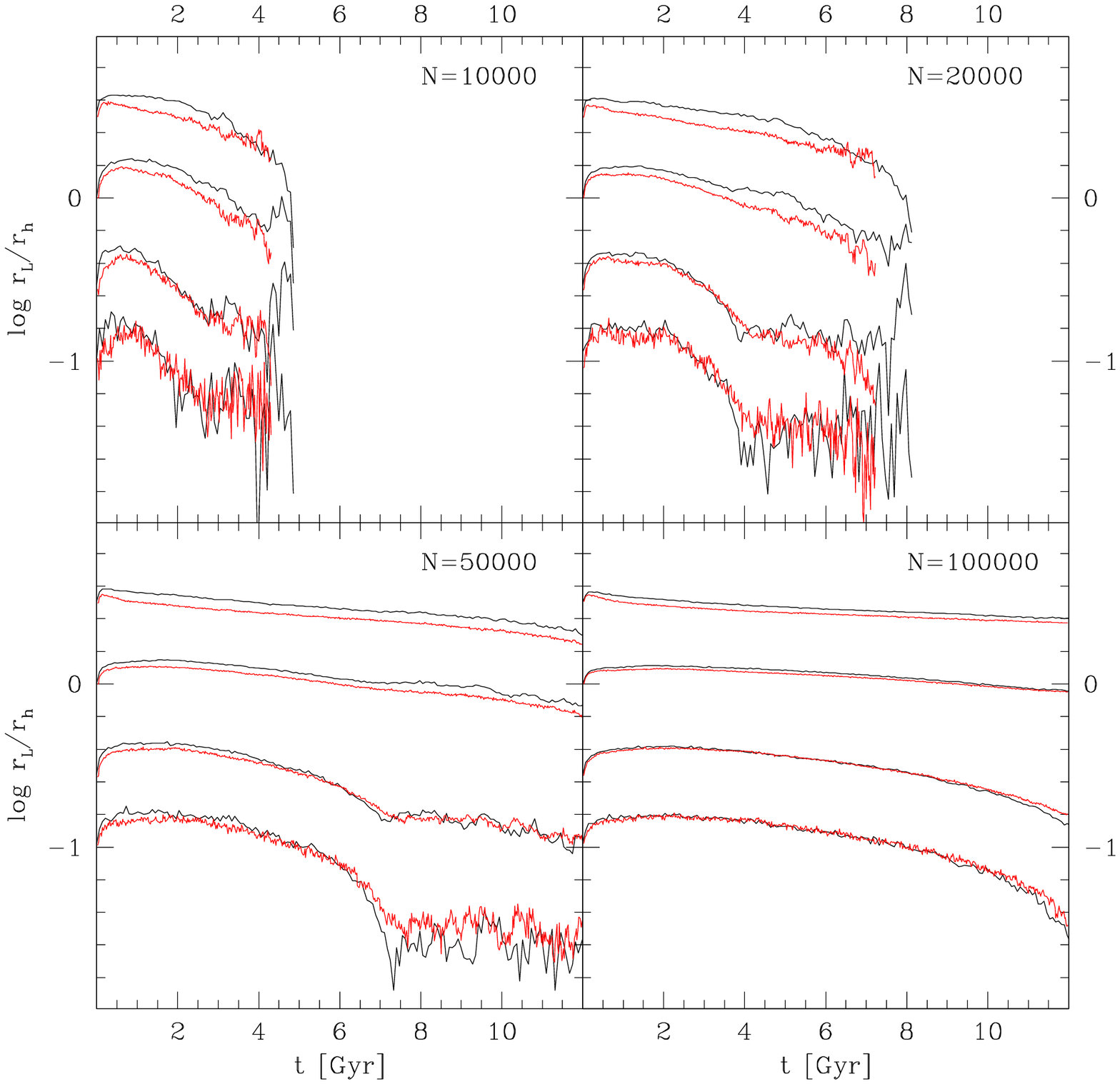}
 \caption{Evolution of the Lagrangian radii containing 1\%, 10\%, 50\% and 90\% of the enclosed mass in the N-body simulations of \citet{baumgardt2003} (black lines) and the corresponding Monte Carlo simulations (red lines).}
\label{mcnb_lag}
\end{figure*}

\begin{figure*}
 \includegraphics[width=\textwidth]{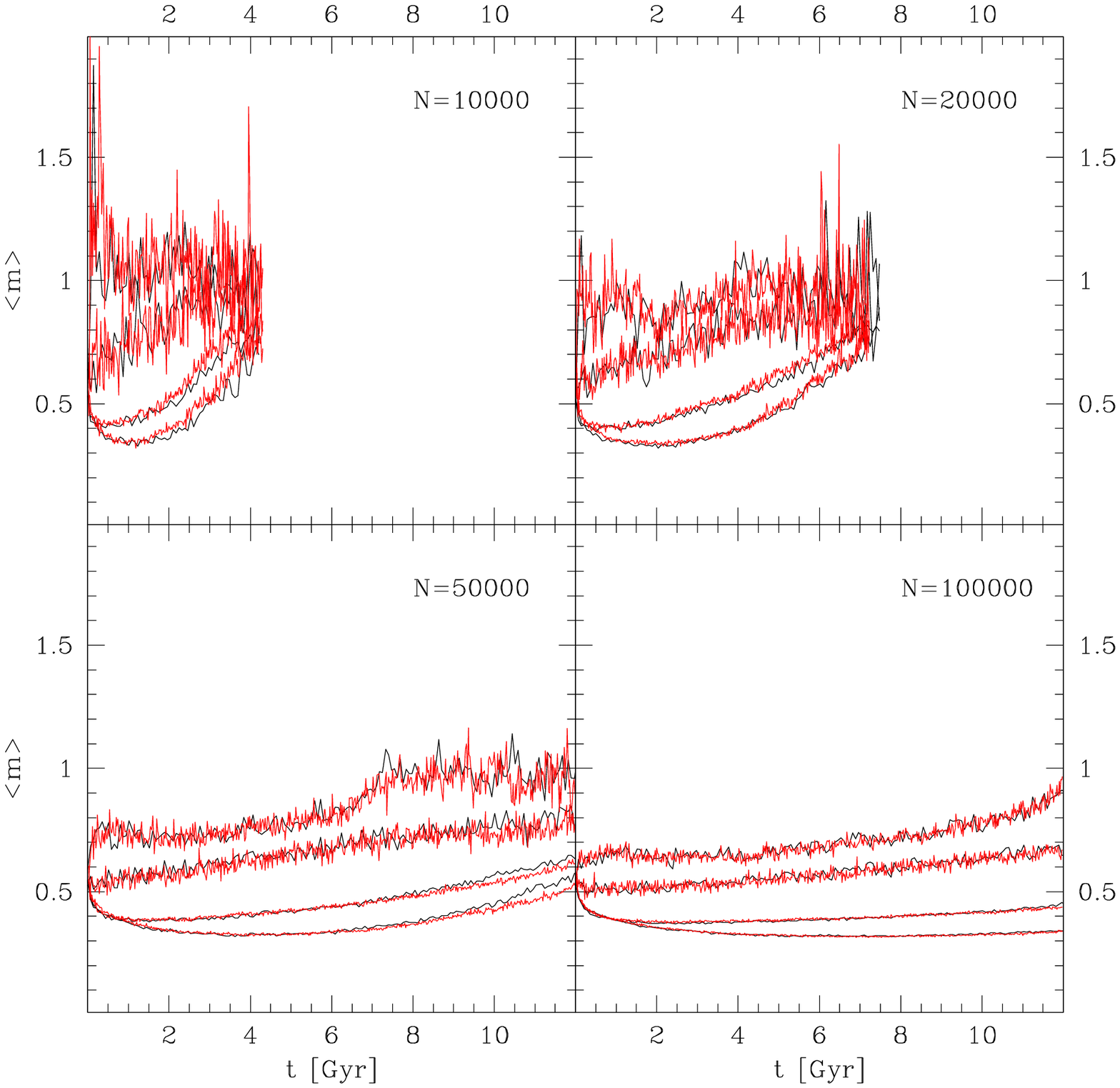}
 \caption{Evolution of the mean mass in the N-body simulations of \citet{baumgardt2003} (black lines) and the corresponding Monte Carlo simulations (red lines).}
\label{mcnb_mm}
\end{figure*}


The basic version of the Monte Carlo code adopted in this paper has been 
extensively tested against N-body simulations in \citet{sollima2014} comparing 
the evolution of the mass and Lagrangian radii of simulations including equal 
mass particles without primordial binaries under different tidal fields.
Since then, many updates have been added to take into account the presence of a mass spectrum, 
stellar evolution, three- and four-bodies interactions \citep{sollima2019,sollima2021}.  
It is therefore necessary to further test the code comparing its predictions with those of N-body simulations including these additional ingredients and/or including multiple populations. 

We consider three different sets of initial conditions for which N-body simulations have been
run by independent groups: {\it i)} the simulations of \citet{vesperini2013} including two populations of equal mass particles without primordial binaries, 
{\it ii)} the simulations of \citet{hong2015} including a single population of equal mass stars with a primordial population of binaries, and 
{\it iii)} the simulations by \citet{baumgardt2003} including a single population with a mass spectrum and stellar evolution.
Each of the above set of simulations adds an additional ingredient to the simple simulations run in \citet{sollima2014}, so that it is 
possible to test the performance of the code in reproducing the effect of each 
process.

The first set of simulations has been presented in \citet{vesperini2013}.
It consists of simulations including 10000 equal mass particles equally distributed between two populations.
Each population is distributed following a \citet{king1966} model with $W_{0}=7$ and different characteristic radii.
We considered simulations with $r_{t,FG}/r_{t,SG}=$10 and 25. No primordial binaries are included in these simulations.
The cluster moves on a circular orbit in a tidal field generated by a point mass truncating the cluster at the tidal radius of the adopted model.
In Fig. \ref{mcnb_vesp} we compare the evolution of the Lagrangian radii of FG/SG stars with the predictions of our Monte Carlo code.
In both simulations, the agreement is excellent, indicating that the dynamical mixing of the two populations is well accounted by our code.

The second set of simulations is taken from the work by \citet{hong2015}.
Among the various simulations run by these authors, we considered those with a single population of 20000 equal mass particles with a fraction of 10\% of primordial binaries.
The particles are distributed following a \citet{king1966} model with $W_{0}=7$. 
The cluster moves on a circular orbit in a tidal field generated by a point mass truncating the cluster at the tidal radius of the adopted model ($R_{J}=r_{t}$).
In each of the considered simulations binaries have been simulated with 
a different hardness parameter $x=E_{b}/m\sigma^{2}$ corresponding to 3 and 20 and following a thermal distribution of eccentricities.
In Fig. \ref{mcnb_hong} the evolution of the fraction of binaries is compared for the two considered simulations. 
It can be seen that in both simulations the declining trend of the number of binaries is well reproduced, with the predictions of the Monte Carlo and N-body codes always agreeing within 10\% along the entire evolution. 
This indicates that the binary formation/destruction is properly accounted by our code.

The third set of simulations has been presented by \citet{baumgardt2003} and consists of a set of simulations including a mass spectrum and a stellar evolution treatment.
Stars are extracted from a \citet{kroupa2001} IMF from 0.1 $M_{\odot}$ to 15 $M_{\odot}$.
We considered the simulations following a \citet{king1966} density profile with $W_{0}=7$ and four different number of particles N=10000, 20000, 50000 and 100000.
Stellar evolution has been implemented following the recipes of \citet{hurley2000}.
No natal kicks have been simulated for massive stars. 
Primordial binaries are not present at the beginning of the simulations but they form during the evolution through tidal and 3-bodies capture.
The cluster moves on a circular orbit in a tidal field generated by a logarithmic potential producing a 
truncation at the tidal radius of the adopted model.
In Fig.s \ref{mcnb_m}, \ref{mcnb_lag} and \ref{mcnb_mm} the evolution of the cluster mass, the Lagrangian radii and the mean mass of the four simulations are compared with the predictions of our Monte Carlo code.
In all cases, the mass, the Lagrangian radii and the mean mass are well reproduced. 
This indicates that the dynamical evolution is well treated by our code also when a spectrum of masses and stellar evolution are included.
A small ($<10\%$) difference is apparent in the mass evolution at the beginning of all simulations, likely resulting from the 
slightly different lifetimes adopted by the two codes. 



\bsp	
\label{lastpage}
\end{document}